\documentclass[12pt]{iopart}
\usepackage{iopams}  

\usepackage{xcolor}
\usepackage{graphicx}

\begin{document}
\title[Bondi accretion]{Bondi accretion in the spherically symmetric Johannsen-Psaltis spacetime}

\author{A J John and C Z Stevens}

\address{Department of Mathematics, Rhodes University, Grahamstown 6139, South Africa}

\ead{a.john@ru.ac.za, c.stevens@ru.ac.za}

\begin{abstract}
The Johannsen--Psaltis spacetime explicitly violates the no hair theorem. It describes rotating black holes with scalar hair in the form of parametric deviations from the Kerr metric.  In principle, black hole solutions in any modified theory of gravity could be written in terms of the Johannsen--Psaltis metric. 
We study the accretion of gas onto a static limit of this spacetime. We utilise a recently proposed pseudo--Newtonian formulation of the dynamics around arbitrary static, spherically symmetric spacetimes. We obtain a potential that generalises the Paczy{\'n}ski--Wiita potential to the static Johannsen--Psaltis metric. We also perform a fully relativistic analysis of the geodesic equations in the static Johannsen--Psaltis spacetime. We find that positive values of the scalar hair parameter, $\epsilon_{3}$, lower the accretion rate and vice versa. Similarly, positive (negative) values of $\epsilon_{3}$ reduce (increase) the gravitational acceleration of radially infalling massive particles.
 \end{abstract}
\submitto{\CQG}
\maketitle

\section{Introduction}

The no--hair theorem \cite{nohair1, nohair2, nohair3} states that black holes in general relativity are uniquely and completely characterised by three parameters viz. their mass, electric charge and angular momentum. An electrically charged black hole will accrete charges of the opposite sign and rapidly neutralise \cite{wald1974black}. Consequently the only physically significant parameters of astrophysical black holes are their mass and angular momentum. The Kerr metric is thus the only stationary, axisymmetric, asymptotically flat vacuum spacetime in general relativity that possesses an event horizon. 

Whilst general relativity has been extremely well constrained on solar system scales \cite{Will2014} the theory has only recently been subjected to tests in strong gravitational fields. The discovery of gravitational waves \cite{abbott2016observation} from the merger of binary black holes provided the first strong field tests of general relativity. Precise determination of the metric of astrophysical black holes allows one to test general relativity in the strong field regime.  Alternative theories of gravity can admit black hole spacetimes that do not comply with the no--hair theorem. Observed violations of the no--hair theorem necessarily imply the breakdown of general relativity.

Given the advent of gravitational wave astronomy as well as forthcoming complementary observations of black holes and neutron stars in the electromagnetic spectrum, strong field tests of gravitational theories are becoming feasible \cite{berti2015testing, broderick2014testing}. A large number of modified theories of gravity have been proposed. Examples of such theories include Modified Gravity (MOG) \cite{moffat2006scalar}, Tensor--Vector--Scalar gravity (TeVeS) \cite{teves} and $f(R)$ gravity \cite{nojiri}. These are motivated by the need to explain the dark matter, dark energy and singularity problems in cosmology as well as the search for a quantum theory of gravity. 

Modified theories of gravity can be tested by solving their associated field equations to obtain black hole solutions and comparing their observable signatures with those predicted by general relativity. Sufficiently precise astronomical data can, in principle, discriminate between competing gravitational theories. Alternatively one could devise a model--independent test of general relativity by constructing black hole spacetimes with parametric deviations from the Kerr solution. These black hole spacetimes are generic in the sense that they are devised without appealing to a specific theory of gravity. Null measurements of these extra parameters would mark a successful test of the no--hair theorem.

The Johannsen--Psaltis (JP) metric \cite{johannsen2011metric} facilitates  model--independent tests of gravitational theories. This metric describes the spacetime of a stationary, axisymmetric black hole that violates the no--hair theorem by construction. Significantly the JP metric does not arise as a solution to field equations of any particular theory of gravity. Given any gravitational theory that violates the no--hair theorem it should be possible to parametrise the deviations from the Kerr metric of its black hole solution(s). In principle these deviations can be related to the additional parameters of the JP metric. The JP metric was not the first proposed generic black hole spacetime. Earlier attempts e.g. \cite{manko1992generalizations} were plagued by problems like the presence of singularities and closed timelike curves outside the event horizon. The JP metric is a promising alternative for parametrising modified gravity corrections to the Kerr solution as it is free of many of these pathologies. 

The JP metric has been applied to a number of problems e.g. strong lensing \cite{jing2012} and the formation of black hole shadows \cite{shadow}. The measurement of black hole spin via continuum fitting and $K\alpha$ iron line methods has also been analysed in the JP spacetime \cite{bambi2013}.

The discovery of active galactic nuclei and quasars as well as compact objects like neutron stars and pulsars prompted the realisation that their high energy emission was due to the liberation of the gravitational potential energy of infalling material \cite{frank2002accretion}. Bondi's seminal paper \cite{bondi1952} determined the mass accretion rate of a polytrope fluid accelerated towards a star with a Newtonian gravitational potential. This result was later extended to full general relativity by Michel \cite{michel1972} who examined accretion onto a Schwarzschild black hole. Accretion disks form when matter falls onto a rotating body. The structure of accretion disks was modelled both in Newtonian gravity \cite{shakura1973} and general relativity \cite{novikov1973}. Black hole accretion in various modified theories of gravity has been investigated by a number of authors \cite{john2013, yang2015, gangopadhyay2018}.

Modelling accreting systems in full general relativity becomes significantly more difficult when one includes more realistic phenomena like viscosity, magnetic fields, turbulence and radiative processes. Paczy{\'n}ski and Wiita \cite{paczynsky1980} introduced a gravitational potential that mimics many features of the Schwarzschild black hole in general relativity. The Paczy{\'n}ski--Wiita potential is commonly used to study black holes in a Newtonian manner \cite{abramowicz2009paczynski}. The advantage of using this pseudo--Newtonian potential is that the equations of motion are substantially simpler to analyse than those arising from a full relativistic treatment. 

Given the successful approximation of black hole physics in general relativity by the Paczy{\'n}ski--Wiita potential, the question arises as to whether pseudo--Newtonian potentials can be found for modified theories of gravity. A number of authors \cite{faraoni2016paczynski,tejeda2014}
have recently devised methods to generate pseudo--Newtonian potentials that could mimic the behaviour of black holes in any static, spherically spacetime. 

In this paper we determine a pseudo--Newtonian potential for the static limit of the Johannsen--Psaltis metric and solve the accretion problem. Our motivation in doing so is to obtain a model--independent formulation of spherically symmetric accretion in modified theories of gravity. In section \ref{sec:jpspace} we introduce the Johannsen--Psaltis metric and its static limit. In section \ref{sec:PWpotential} we obtain a pseudo--Newtonian potential for the JP metric. In Section \ref{sec:accretion} we solve the accretion problem for JP black holes approximated by our potential. In section \ref{sec:geodesics} we analyse the motion of test particles in the JP spacetime. Our results are summarised in Section \ref{sec:summary}. Throughout this paper, unless explicitly stated, we utilise geometric units where $c=1=G$.

\section{The Johannsen-Psaltis spacetime}\label{sec:jpspace}

The Johannsen-Psaltis metric takes the form
\begin{eqnarray}
	ds^2 = &-[1+h(r,\theta)]\Big{(}1-\frac{2mr}{\Sigma}\Big{)}dt^2 
		    - \frac{4amr\sin^2\theta}{\Sigma}[1+h(r,\theta)]dtd\phi\nonumber \\
		   &+ \frac{\Sigma[1 + h(r,\theta)]}{\Delta + a^2\sin^2\theta h(r,\theta)}dr^2
		    + \Sigma d\theta^2 \nonumber \\
		   &+ \Big{[}\sin^2\theta\Big{(}r^2 + a^2 
		    + \frac{2a^2mr\sin^2\theta}{\Sigma}\Big{)} \nonumber \\ 
		   &+ h(r,\theta)\frac{a^2(\Sigma + 2mr)\sin^4\theta}{\Sigma}\Big{]}d\phi^2,
\end{eqnarray}
where
\begin{eqnarray}
	\Sigma := r^2 + a^2\cos^2\theta, \\
	\Delta := r^2 - 2mr + a^2,
\end{eqnarray}
and $m$ and $a$ are the black hole mass and spin parameters respectively. Here $h(r,\theta)$ is the parametric deviation from the Kerr spacetime
\begin{equation}
	h(r,\theta) := \sum_{k=0}^{\infty}\Big{(}\epsilon_{2k} 
				   + \epsilon_{2k+1}\frac{mr}{\Sigma}\Big{)}\Big{(}\frac{m^2}{\Sigma}\Big{)}^k
\end{equation}
and the $\epsilon$ terms are dimensionless coefficients.
It can easily be seen that when $h(r,\theta)=0$ this reduces to the standard Kerr spacetime.
We are interested only in the spherically symmetric case, which reduces the form of the metric to
\begin{eqnarray}\label{eq:SSJPmetric}
	ds^2= &- [1+h(r)]\Big{(}1 - \frac{2m}{r}\Big{)}dt^2 
		  + [1+h(r)]\Big{(}1 - \frac{2m}{r}\Big{)}^{-1}dr^2 \nonumber \\
		  &+ r^2\Big{(}d\theta^2 + \sin^2\theta d\phi^2\Big{)},
\end{eqnarray}
where the parametric deviation now takes the form
\begin{equation}\label{eq:SSdeviationh}
	h(r) = \sum_{k=0}^{\infty}\Big{(}\epsilon_{2k} 
				   + \epsilon_{2k+1}\frac{m}{r}\Big{)}\Big{(}\frac{m}{r}\Big{)}^{2k}.
\end{equation}
In the absence of any scalar hair i.e. $\epsilon_{k} = 0$ we recover the line element for the Schwarzschild black hole. 

There are a number of observational constraints on the magnitude of the $\epsilon_{k}$ parameters \cite{johannsen2011metric}. In general relativity stationary and asymptotically flat spacetimes must fall off as $1/r$ or faster. Spacetimes with slower fall off rates will involve gravitational radiation and thus cannot be stationary. Similar arguments hold for spacetimes that don't arise as solutions of the Einstein field equations. Thus the function $h(r)$ must be of order $\mathcal{O}(r^{-n})$ where $n \geq 2$ and this forces $\epsilon_{0} = \epsilon_{1} = 0$.
Weak field deviations from general relativity can be determined using the parameterized post--Newtonian (PPN) framework. The Lunar Laser Ranging experiment \cite{Will2014} sets an upper bound of $|\epsilon_{2}| \leq 4.6 \times 10^{-4}$. The first unconstrained parameter in $h(r)$ is thus $\epsilon_{3}$. Analyses involving the JP spacetime often only consider the leading non--vanishing contribution to $h(r)$ i.e. the scalar hair is approximated via
\begin{equation}
h(r) = \epsilon_3\Big{(}\frac{m}{r}\Big{)}^3. \label{eq:truncated}
\end{equation}

Throughout this paper we will either use the full expression for $h(r)$ viz. 
\eref{eq:SSdeviationh} 
or the truncated version,
\eref{eq:truncated}.

\section{The Paczy{\'n}ski-Wiita-like potential}
\label{sec:PWpotential}

In this section we derive the Paczy{\'n}ski--Wiita like potential $\Phi_{JP}$ for the Johannsen-Psaltis metric.

The original Paczy{\'n}ski--Wiita potential is \emph{pseudo-Newtonian} in the sense that it is a potential representing the relativistic Schwarzschild spacetime that can be used in a Newtonian framework. The potential is
\begin{equation}\label{eq:PWpotential}
	\Phi_{PW} := -\frac{m}{r-2m},
\end{equation}
where $m$ is the mass parameter from the Schwarzschild metric.

Although the form of this potential was essentially ``guessed'' by Paczy{\'n}ski, there has subsequently been a method found which when applied to the Schwarzschild spacetime gives the same potential \cite{abramowicz2009paczynski}. This method has been applied to general static spherically symmetric space-times in \cite{faraoni2016paczynski} which we now summarise.

Any static spherically symmetric spacetime written in coordinates $\{t,r,\theta,\phi\}$ can be put into the form
\begin{equation}
	ds^2 = g_{00}(r)dt^2 + g_{11}(r)dr^2 + r^2d\Omega^2,
\end{equation}
where $d\Omega^2=d\theta^2+\sin^2\theta d\phi^2$.  
The derivation of the pseudo--Newtonian potential relies on examining the orbit of a test particle in the spacetime, which without loss of generality can be restricted to the $\theta=\pi/2$ plane. The effective gravitational potential can be identified and takes the form
\begin{equation}
	\Phi := \frac{1}{2}(g^{00}(r) + k),
\end{equation}
where $k\in\mathbb{R}$ is an arbitrary constant. When applied to the Schwarzschild spacetime, the choice $k=1$ yields the original Paczy{\'n}ski-Wiita potential.

Turning our attention to the spherically-symmetric Johannsen-Psaltis metric Eq. \eref{eq:SSJPmetric}, we find the corresponding potential when again choosing $k=1$ is
\begin{equation}
	\Phi_{JP} := -\frac{m}{r-2m} + \frac12\frac{rh(r)}{[1 + h(r)](r-2m)}, \label{phiJP}
\end{equation}
which clearly reduces to $\Phi_{PW}$ when $h(r)=0$. Note that this expression uses the full expansion of $h(r)$ given by \eref{eq:SSdeviationh}.

\section{Bondi accretion}\label{sec:accretion}
We now compute the mass accretion rate of a gas falling into a black hole described by the line element \eref{eq:SSJPmetric}. In the spirit of Paczy{\'n}ski and Wiita we model this problem using classical hydrodynamics and the effective gravitational potential, $\Phi_{JP}$. 

We model the gas as a perfect fluid with a polytrope equation of state, $p = K \rho^{\gamma}$, where the adiabatic index, $\gamma$, is given by the ratio of specific heats i.e. $\gamma = c_{p} / c_{v}$.

The dynamical motion of the fluid is determined by the conservation of mass and momentum viz.
\begin{eqnarray}
	\frac{\partial\rho}{\partial t} + \nabla\cdot(\rho \mathbf{u}) = 0 \\
	\rho \left[\frac{\partial \mathbf{u}}{\partial t} + ( \mathbf{u} \cdot \nabla)\mathbf{u}\right] 
	= -\nabla p - \rho \nabla \Phi_{JP},
\end{eqnarray}
where $\mathbf{u}$ is the fluid 4-velocity, $p$ is the isotropic pressure and $\rho$ is the mass density. The effective pseudo--Newtonian potential $\Phi_{JP}$ is given by \eref{phiJP}.

Imposing spherical symmetry, and considering only radially infalling steady-state flows  gives us the simpler system
\begin{eqnarray}
	\frac{1}{r^2}\frac{d}{dr}(\rho u r^2) = 0,\label{eq:continuity} \\
	u\frac{du}{dr} + \frac{1}{\rho}\frac{d\rho}{dr}+\frac{d\Phi_{JP}}{dr} = 0,\label{eq:mattermotion} 
\end{eqnarray}
where $u(r)$ is the radial velocity of the fluid. For convenience we rewrite the equation of state using the polytropic index, $\displaystyle p = K \rho^{1 + \frac{1}{n}}$.
Now we can integrate Eqs. \eref{eq:continuity} and \eref{eq:mattermotion} to find the conserved mass flux and specific energy respectively as
\begin{eqnarray}
	\dot{M}     = \rho u r^2,\label{eq:Mdot}	
\\
	\mathcal{E} = \frac12u^2 + na^2 + \Phi_{JP} = na_\infty^2, \label{eq:energy}
\end{eqnarray}
where we have denoted $\displaystyle a=\sqrt{\frac{d p}{d \rho}}$ as the adiabatic sound speed. 
Our boundary conditions are determined at a point, $r_{\infty}$, which is sufficiently far away from the black hole that the fluid velocity ($u(r_{\infty}) =: u_{\infty}$) and the gravitational potential both vanish. The fluid sound speed ``at infinity'' is denoted by $a_\infty$.
 Further, we rewrite \Eref{eq:Mdot} to obtain the accretion rate
\begin{equation}\label{eq:accretion}
	\dot\mathcal{M} = a^{2n}ur^2,
\end{equation}
which is related to the mass flux via $\dot\mathcal{M} = \dot M\gamma^nK^n$ \cite{chakrabarti1990theory}. 
We now combine Eqs. \eref{eq:energy} and \eref{eq:accretion} to obtain
\begin{equation}
	\frac{du}{dr} = \frac{\displaystyle\frac{2a^2}{r} 
					- \frac{d\Phi_{JP}}{dr}}{\displaystyle u - \frac{a^2}{u}}. \label{wind}
\end{equation}

To ensure that the accretion flow is smooth, the denominator of \eref{wind} must vanish at the same point as the numerator. This \emph{critical point}, $r_{c}$ is determined by the conditions:
\begin{equation}\label{eq:criticalpconds}
	a_c = u_c,\qquad \frac{d\Phi_{JP}}{dr}\bigg |_{r=r_c} = \frac{2a_c^2}{r_c}.
\end{equation}
The first condition tells us that the critical point is also a \emph{sonic point}. If the fluid accelerates from rest  far away from the black hole ($u_{\infty}=0$) it must reach its local sound speed at the critical point i.e. $u_{c} = a_{c}$.

Solving \eref{eq:criticalpconds} for the critical points is equivalent to finding the roots of the equation
\begin{equation}\label{eq:infinitepoly}
	2[m r_c-2a_c^2(r_c-2m)^2]+r_{c}h(r_c)(4m-3r_c).
\end{equation}
This equation is valid for the full form of $h(r)$ given by \eref{eq:SSdeviationh}. This expression is quite complicated in general so we only consider the leading contribution to $h(r)$ given by \eref{eq:truncated}, which is commonly done in studies of the JP spacetime \cite{johannsen2011metric,zelenka2017chaotic}.

The specialisation
\begin{equation}
	h_{c}(r_{c}) = \epsilon_3\Big{(}\frac{m}{r_{c}}\Big{)}^3.
\end{equation}
 reduces Eq. \eref{eq:infinitepoly} to a quartic polynomial in $\alpha := r_{c}/m$ which has the form
\begin{equation} \label{eq:quartic}
	-2\epsilon_3 + \frac{3}{2}\epsilon_3\alpha + 8a_c^2\alpha^2 - (1 + 8a_{c}^{2})\alpha^3 + 2a_c^2\alpha^4,
\end{equation}
where we must compute the roots to relate $r_c$ to $a_c$. It is not surprising that even in the simplified case of $h(r)$ considered, the four roots of this quartic are lengthy expressions. We observed that for a wide range of values for $\epsilon_3$ and $a_{c}$ only one of the roots is always real and outside the event horizon located at $r=2m$\footnote{The location of the event horizon in the static limit of the Johannsen-Psaltis metric is still at $r=2m$.}. We identify this root as \emph{the} critical point of the accretion flow.

To obtain a more tractable analytical expression for this root, we linearise \eref{eq:quartic} about $\epsilon_3$ for the rest of this section. 

The physically relevant root is given by
\begin{equation}\label{eq:rootofpoly}
	\alpha_c = \frac{1 + 8a_{c}^{2} + X}{4a_{c}^{2}} 
		  + \frac{1 - X - 2a_{c}^{2}\Big{(}-5 + 8a_{c}^{2} + X\Big{)}}{64a_{c}^{4} X}\epsilon_3 + O(\epsilon_3)^{3/2},
\end{equation}
with $X:=\sqrt{1+16a^2}$. This is further simplified by taking our equations to be linear in $a_{c}^{2}$ as done in \cite{shapiro2008black}. This gives the simple expression
\begin{equation}\label{eq:criticalpointrfinal}
	r_c \approx \left( \frac{1 + (8 - \epsilon_{3}) a_{c}^{2}}{ 2 a_{c}^{2}} \right) m,
\end{equation}
which reduces to the well known result for the Schwarzschild spacetime found in the literature \cite{michel1972, shapiro2008black}.

Using Eq. \eref{eq:criticalpointrfinal} and Eq. \eref{eq:energy} and linearising about $\epsilon_3$, $a_{c}^{2}$ and $a^2$ we find the relationship between the critical sound speed and the boundary condition as
\begin{equation}
	a_c^2 = \frac{2n}{2n-3} a_{\infty}^{2},
\end{equation}
which is the same as the Schwarzschild case.

Using the above together with Eq. \eref{eq:accretion} we find the Bondi accretion rate,
\begin{equation}
\dot{\mathcal{M}} = \frac{1}{4} \left( \frac{2n}{2n-3}  a_{\infty}^{2} \right)^{(2n-3) / 2} \left( 1 + \frac{ 4 (8 - \epsilon_{3}) n a_{\infty}^{2}}{2n-3} \right) m^{2} \label{eq:mdot}
\end{equation}
which is kept in this form (i.e. not being completely linear in $a_\infty^2$) for simplicity.

\subsection{Comparison with the Schwarzschild case}{}
To compare pseudo--Newtonian accretion in the Johannsen--Psaltis and Schwarzschild spacetimes  we investigate a range of different system parameters and tabulate how $\epsilon_3$ affects the accretion rate. These parameters are $m, \gamma \Leftrightarrow n$ and $a_\infty$. For the below we fix the boundary condition $a_\infty=3\times10^{-6}$ which follows from the temperature of ionised interstellar gas  \cite{john2013}. We consider two fiducial values for the black hole mass viz. a solar mass black hole, $m=1 M_{\odot}$, and a supermassive black hole, $m=10^{6} M_{\odot}$. We restrict $\gamma$ to have values $1<\gamma<5/3$. We write $\dot\mathcal{M}=A+B\epsilon_3$ in order to tabulate the results.

\begin{table}
\caption{\label{AccretionRateSolarMass} The  accretion rate of matter surrounding a $1 M_{\odot}$ black hole}
\begin{tabular*}{\textwidth}{@{}l*{15}{@{\extracolsep{0pt plus
12pt}}l}}
\br
$\gamma$&n&A&B\\
\mr
1.1&10&$1.67654\times10^{-70}$&$-4.38311\times10^{-79}$\\
1.2&5&$8.66651\times10^{-26}$&$-2.75127\times10^{-34}$\\
1.3&10/3&$6.24348\times10^{-11}$&$-2.52262\times10^{-19}$\\
4/3&3&$5.69793\times10^{-8}$&$-2.53241\times10^{-16}$\\
1.4&2.5&$1.51089\times10^{-3}$&$-8.39385\times10^{-12}$\\
1.5&2&$3.62614\times10^{1}$&$-3.22324\times10^{-7}$\\
\br
\end{tabular*}
\end{table}

\begin{table}
\caption{\label{AccretionRateStellarMass} The  accretion rate of matter surrounding a $10^{6} M_{\odot}$ black hole}
\begin{tabular*}{\textwidth}{@{}l*{15}{@{\extracolsep{0pt plus
12pt}}l}}
\br
$\gamma$&n&A&B\\
\mr
1.1&10&$1.67654\times10^{-58}$&$-4.38311\times10^{-67}$\\
1.2&5&$8.66651\times10^{-14}$&$-2.75127\times10^{-22}$\\
1.3&10/3&$6.24348\times10^1$&$-2.52262\times10^{-7}$\\
4/3&3&$5.69793\times10^4$&$-2.53241\times10^{-4}$\\
1.4&2.5&$1.51089\times10^9$&$-8.39385\times10^0$\\
1.5&2&$3.62614\times10^{13}$&$-3.22324\times10^{2}$\\
\br
\end{tabular*}
\end{table}

One sees from the Tables \eref{AccretionRateSolarMass} and \eref{AccretionRateStellarMass} that $\epsilon_3$ affects the accretion rate $A$ predicted by general relativity by adding the number $B$ which, if $A$ has order $a$, then $B$ has order roughly $a-8+e$, where $e$ is the order of $\epsilon_3$. Note that we took $\epsilon_3$ to be small to arrive at Eq. \eref{eq:rootofpoly}, and thus $e$ should mimic this restriction. When $\epsilon_3>0$ the accretion rate is lowered, while when $\epsilon_3<0$ the accretion rate is raised. This behaviour is consistent with a back--of--the--envelope calculation of the gravitational force of our system. For our potential, $\Phi_{JP}$, we have $F \sim - \Phi_{JP}' \sim -\frac{1}{r^{2}} + \epsilon_{3} \frac{1}{r^{4}}$. For positive values of $\epsilon_{3}$ the gravitational force becomes less negative and hence less attractive. Similarly negative values of $\epsilon_{3}$ result in a more negative and thus larger attractive force. Further, when $\gamma$ increases, so does the accretion rate. This is consistent with other results in the literature \cite{john2016black}.

\section{Test particle orbits} \label{sec:geodesics}
To further investigate the accretion properties of the previous section, we perform a fully relativistic calculation of test particle orbits in the spherically-symmetric Johannsen-Psaltis spacetime and how they are effected by a non-zero $\epsilon_3$. Note that we do not linearise about $\epsilon_3$ in this section.

First, we obtain the constants of motion of a test particle travelling in the $\theta=\pi/2$ plane with 4-velocity $u^a = \{t'(\lambda),r'(\lambda),0,\phi'(\lambda)\}$ parametrised by $\lambda$ and satisfying $u^au_a=-{\kappa}^2$:
\begin{eqnarray}
	E = \Big{(}1-\frac{2m}{r(\lambda)}\Big{)}\label{eq:conservedenergy}
	    \Big{(}1 + \frac{\epsilon_3m^3}{r(\lambda)^3}\Big{)}t'(\lambda), \\
	L = r(\lambda)^2\phi'(\lambda), \\
	\frac{E^2}{2} = V(\lambda) + \frac12\Big{(}1+\frac{\epsilon_3m^3}{r(\lambda)^3}\Big{)}r'(\lambda)^{2}, \label{eq:radialequation}\\
	V(\lambda) \equiv \frac{\kappa}{2} - \frac{\kappa m}{r(\lambda)} + \frac{L^2}{2r(\lambda)^2}
				 - \frac{L^2m}{r(\lambda)^3} \nonumber \\
	+ \epsilon_3\Big{(}\frac{\kappa m^3}{2r(\lambda)^3}
	- \frac{\kappa m^4}{r(\lambda)^4} + \frac{L^2m^3}{2r(\lambda)^5}
	- \frac{L^2m^4}{r(\lambda)^6}\Big{)}.\label{eq:potential}
\end{eqnarray}
Here $E$ and $L$ are the conserved energy and angular momentum, $V(\lambda)$ is the potential and $\kappa=1$, $\kappa=0$ correspond to timelike and null orbits respectively.

To get an idea of how this timelike test particle behaves outside the event horizon with differing choices of $\epsilon_3$, we can numerically solve Equation \eref{eq:radialequation} with potential given by Equation \eref{eq:potential} by choosing
\begin{equation}
	m = M_\odot\approx1475, \quad L = 0.1, \quad E=0.5, \quad \kappa=1, \quad r(0)=3500.
\end{equation}
\begin{figure}
	\begin{center}
		\includegraphics[width=0.6\linewidth]{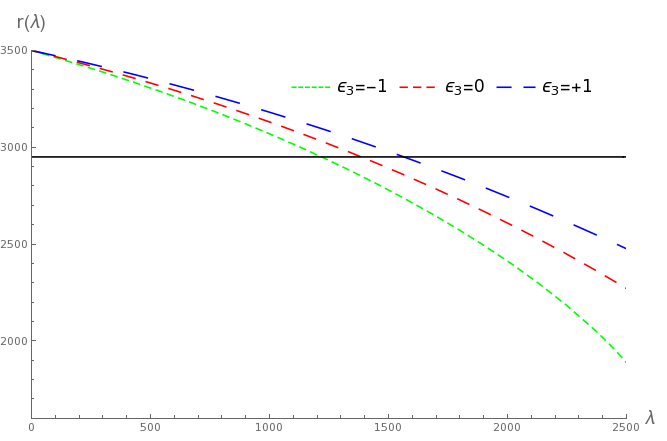}
	\end{center}
	\caption{\label{fig:JPtrajectories}Timelike test particle trajectories with varying $\epsilon_3$. The solid horizontal line indicates the event horizon.}
\end{figure}
The results are plotted in Figure \eref{fig:JPtrajectories} for different values of the $\epsilon_3$ parameter\footnote{These parameters are not valid in the linear regime but are taken for visualisation purposes.}. One can see that $\epsilon_3>0$ has the effect of slowing down the particles approach toward the central object. This is in direct agreement with the slower accretion rate calculated in Section \eref{sec:accretion}. For the choice $\epsilon_3<0$ the opposite is true and again matches with a faster accretion rate.

It is useful to present the linearised (in $\epsilon_3$) unstable circular photon orbit
\begin{equation}
	r = 3m - \epsilon_3\frac{m}{18}, 	
\end{equation}
which is found by solving $\displaystyle\frac{dV}{dr}=0$ and is in agreement with \cite{johannsen2011metric} once linearised and appropriate limits taken. This simple expression shows that with $\epsilon_3>0$ and $|\epsilon_3| \ll 1$ the photon orbit is slightly closer to the black hole. This again supports the paradigm of a slower accretion rate.
We note that the Innermost Stable Circular Orbit (ISCO) does not change from $r=6m$, even without linearising $\epsilon_3$.
The fully relativistic analysis of geodesic motion supports the results of our pseudo--Newtonian calculation of the accretion rate, $\dot{\mathcal{M}}$ as well as the back--of--the--envelope calculation in the previous section. 

\section{Summary} \label{sec:summary}

We examined the problem of accretion onto a static black hole that generically violates the no hair theorem. The Johannsen--Psaltis spacetime describes a rotating black hole with scalar hair and does not arise from a specific theory of gravity. We restricted our focus to the static limit of the JP metric. This limit contains parametric deviations from the Schwarzschild solution of general relativity.

We studied a Newtonian formulation of the accretion problem using the method of Faraoni \emph{et al} \cite{faraoni2016paczynski}. This technique generalises the Paczy{\'n}ski--Wiita potential to general static spherically symmetric black hole spacetimes. We obtained a pseudo--Newtonian potential that approximates many of the gravitational features of the static JP metric.

We modelled the matter accreting onto a static JP black hole as a polytropic fluid. The fluid is at rest far from the black hole and accelerates radially inwards at subsonic speeds. The fluid's  speed reaches its local sound speed at a critical point. Thereafter the fluid continues to accelerate towards the event horizon supersonically. 

In order to determine the critical point we solved a fourth order polynomial equation numerically. Only one of the four roots was found to be real, positive and outside the event horizon. This root was identified as \emph{the} critical point of the transonic flow. We linearised the quartic equation about $\epsilon_{3}$ and obtained an approximate analytical expression for the critical point's position. We then obtained an analytical expression for the critical velocity of the fluid in terms of its sound speed at ``infinity''. The critical velocity is independent of $\epsilon_{3}$.

We obtained an analytical expression for the accretion rate, $\dot{\mathcal{M}}$. Our accretion rate is proportional to the square of the black hole mass i.e. $\dot{\mathcal{M}} \sim m^{2}$. This is a common feature in Bondi accretion. The accretion rate depends on the black hole mass, $m$, the fluid's polytrope index, $n$, the fluid sound speed at infinity and the JP parameter, $\epsilon_{3}$.
Positive(negative) values of $\epsilon_{3}$ were found to reduce(increase) the accretion rate and the accretion rate increased with $\gamma$. These results were tabulated.

We obtained and solved the geodesic equations for massive and massless particles orbiting a static JP black hole. Positive values of $\epsilon_{3}$ were found to slow down massive particles in radially infalling trajectories around a static JP black hole. Similarly, negative values of $\epsilon_{3}$ accelerated massive particles. Our fully relativistic analysis of geodesic trajectories corroborates our findings on the effect of $\epsilon_{3}$ on the accretion rate viz. that small, positive values of $\epsilon_{3}$ lowered the accretion rate and vice versa.

There is no a priori reason to expect an overall positive or negative sign to scalar hair contributions to a black hole metric. Black holes with scalar hair arising in modified theories of gravity will thus either increase or decrease the efficiency of the conversion of gravitational energy into radiant energy. Our results can be used to vindicate the validity of the pseudo--Newtonian approach to black holes. The problem we investigated was an idealised study of transonic accretion onto a JP black hole. This framework can used to incorporate various important physical phenomena like radiative processes, viscous dissipation, magnetic fields and accretion disks.

\ack

AJJ and CZS thank Rhodes University for financial support.

\section*{References}

\bibliographystyle{elsarticle-num} 
\bibliography{refs2}

\end{document}